\documentclass[12pt,leqno]{amsart}
\usepackage{graphics}

\newcommand{\bfr}{\displaystyle\frac}

\newcommand{\ep}{\varepsilon}
\newtheorem{theorem}{Theorem}

\newcommand{\Bb}{\mathbb}
\newtheorem{proposition}{Proposition}
\newtheorem{definition}{Definition}
\theoremstyle{remark}
\newtheorem{remark}{Remark}

\begin{document}

\title[Encryption Using Frames]{Encryption Schemes using Finite Frames and Hadamard Arrays}
\author{Ryan Harkins}
\address{Department of Mathematics, University of Wyoming, Laramie, WY 82071-3036}
\email{narandus@hotmail.com}

\author{Eric Weber}
\address{Department of Mathematics, University of Wyoming, Laramie, WY 82071-3036}
\curraddr{Deparment of Mathematics, Iowa State University, 400 Carver Hall, Ames, IA 50011}
\email{esweber@iastate.edu}

\author{Andrew Westmeyer}
\address{Department of Mathematics, University of Wyoming, Laramie, WY 82071-3036}
\email{westandy@uwyo.edu}

\date{\today}

\begin{abstract}
We propose a cipher similar to the One Time Pad and McEliece cipher based on a subband coding scheme.  The encoding process is an approximation to the One Time Pad encryption scheme.  We present results of numerical experiments which suggest that a brute force attack to the proposed scheme does not result in all possible plaintexts, as the One Time Pad does, but still the brute force attack does not compromise the system.  However, we demonstrate that the cipher is vulnerable to a chosen-plaintext attack.
\end{abstract}

\maketitle

\section{Introduction}


In this paper, we propose a private key cipher, the idea for which comes from frame theory and multiple access communications.  The cipher has similarities to the Hill cipher, the One Time Pad, and the McEliece cipher \cite{HAC,Cha}.  Indeed, one of the design goals for our cipher is to approximate the One Time Pad.

Our design goals include the following:
\begin{enumerate}
\item Include randomness in the encryption process;
\item Require the key be shared only once;
\item Use a relatively small key size;
\item Computationally fast;
\item Robust to brute force attacks.
\end{enumerate}
Our proposed cipher implements items 1-4 above; the purpose of the present paper is to give some demonstration of item 5.  We remark here that 5 is not sufficient for the cipher to be a good one, but certainly is necessary.  We will demonstrate that this cipher is vulnerable to a chosen-plaintext attack.  It is unknown if this cipher is robust against a known-plaintext attack.

Our cipher can be described as follows:  consider a communications channel; we divide the channel into two subbands, one which will carry the message, and the other which will carry noise, or as we call it in this paper, garbage.  The message, along with the garbage is transmitted over the channel; the recipient then filters out the garbage, leaving only the message.  This procedure is carried out using orthogonal frames.  The procedure requires the construction of orthogonal frames; the easiest way to do this is using Fourier frames (also called harmonic frames).  However, as will be described, these frames are not good for our purposes here, and so we present several alternative methods for constructing orthogonal frames.

The paper is organized as follows.  In Section \ref{S:frames}, we give a short introduction to frames, and in particular orthogonal frames.  In Section \ref{S:schemes}, we give an account of several methods for constructing orthogonal frames, with remarks regarding our design goals.  In Section \ref{S:results}, we present the results and conclusions of our numerical experiments and the chosen-plaintext attack.  In the Appendix, we provide psuedocode to describe the experiments.



\section{Introduction to Frames} \label{S:frames}

Frames for Hilbert spaces are being used in many signal processing applications such as sampling theory, multiple access communications, etc.  Frames provide redundancy via overcompleteness, where bases do not, and it is this redundancy that makes them advantageous to use in these settings.  In this paper, we will utilize this redundancy of frames for the purpose of encryption.

Let $H$ be a Hilbert space over the field $\Bb{F}$ with scalar product $\langle \cdot , \cdot \rangle$ and norm $\| \cdot \|$, where $\Bb{F}$ denotes either $\Bb{R}$ or $\Bb{C}$.  A frame for $H$ is a sequence $\Bb{X} := \{x_n\}_{n \in \Bb{Z}}$ such that there exist constants $0 < A \leq B < \infty$ such that for all $v \in H$,
\begin{equation} \label{E:frame}
A \|v \|^2 \leq \sum_{n \in \Bb{Z}} | \langle v, x_n \rangle |^2 \leq B \|v \|^2.
\end{equation}
Clearly, a frame spans the Hilbert space.  Moreover, $\{x_n\}$ defines the following \emph{frame operator}
$$ S_{\Bb{X}}: H \to H: v \mapsto \sum_{n \in \Bb{Z}} \langle v , x_n \rangle x_n $$
which is positive and invertible.  Define $\{\tilde{x}_n\} \subset H$, the \emph{standard dual} of $\{x_n\}$ by $\tilde{x}_n := S_{\Bb{X}}^{-1} x_n$, then for all $v \in H$,
$$ v = \sum_{n \in \Bb{Z}} \langle v , x_n \rangle \tilde{x}_n = \sum_{n \in \Bb{Z}} \langle v , \tilde{x}_n \rangle x_n.$$
If $A = B = 1$, the frame is said to be \emph{Parseval}, and then for all $v \in H$,
$$ v = \sum_{n \in \Bb{Z}} \langle v , x_n \rangle x_n. $$
For elementary frame theory, see \cite{HL, Casazza}.

If $H$ is finite dimensional ($H$ will always be assumed to be so from here on, unless specifically stated), then a frame sequence (possibly finite) is any spanning set $\{x_n\}$ such that $\sum_{n \in \Bb{Z}} \| x_n \|^2 < \infty$.  If only a finite number of $x_n$'s are non-zero, then $\{x_n\}$ is a finite frame, and we will discard those that are zero.  See \cite{Casazza2, Dykema, Fickus} for more on finite frames.

For convenience of notation, we make the following definition.
\begin{definition} An $n\times n$ real matrix, $M$, is an orthogonal matrix if $M^TM=kI_n$ for some constant $k$.  \end{definition}

The (finite) Parseval frames in $H$ are characterized by the following proposition.
\begin{proposition} \label{P:PF}
Let $\{x_n\}_{n=1}^{M} \subset H$, where $H$ has dimension $N$.  The following are equivalent:
\begin{enumerate}
\item $\{x_n\}$ is a Parseval frame for $H$;
\item the $M \times N$ matrix whose $i$th row is $x_i$ (as a row vector) has columns which are orthonormal;
\item there exists a Hilbert space $K$ of dimension $M-N$ and vectors $\{y_n\}_{n =1}^{M} \subset K$ such that the $M \times M$ matrix formed by
$$
\left(
\begin{aligned}
&x_1 & &| & &y_1 \\
& \vdots & &| & & \vdots \\
&x_M & &| & &y_M
\end{aligned}
\right)
$$
is a unitary matrix.
\end{enumerate}
\end{proposition}

Here we write the vectors $x_i$ and $y_i$ as row vectors with respect to any orthonormal bases for $H$ and $K$, respectively.
\begin{proof}
The proof of the equivalence of 1 and 2 is in \cite{Fickus}.  The proof of the equivalence of 1 and 2 is, for infinite frames, contained in \cite[Corollary 1.3, Theorem 1.7]{HL}.  The case for finite frames is analogous.
\end{proof}

\begin{remark}
Another way to view Proposition \ref{P:PF} is that $\{x_n\}$ is a Parseval frame for $H$ if and only if $\{x_n\}$ is the inner direct summand of an orthonormal basis $\{x_n \oplus y_n\}$ for some superspace $H \oplus K$ of $H$.
\end{remark}

\begin{definition}
Two frames $\{x_n\}_{n=1}^{M} \subset H$ and $\{y_n\}_{n=1}^{M} \subset K$ are \emph{orthogonal} if for all $v \in H$, $\sum_{n=1}^{M} \langle v , x_n \rangle y_n = 0$.
\end{definition}

\begin{proposition} \label{P:ortho}
Suppose $\{x_n\}_{n=1}^{M} \subset H$ and $\{y_n\}_{n=1}^{M} \subset K$ are Parseval frames; they are orthogonal if and only if
$$
\left(
\begin{aligned}
&x_1 & &| & &y_1 \\
& \vdots & &| & & \vdots \\
&x_M & &| & &y_M
\end{aligned}
\right)
:= \left(P | Q \right)
$$
has columns which form an orthonormal set.
\end{proposition}
\begin{proof}
($\Leftarrow$) Consider the two matrices $P$ and $Q$ whose rows are $\{x_n\}$ and $\{y_n\}$, respectively.  A straight forward computation demonstrates that for $v \in H$,
\begin{equation} \label{E:ortho}
\sum_{n =1}^{M} \langle v , x_n \rangle y_n = Q^{*} P v,
\end{equation}
where $Q^{*}$ is the conjugate transpose of $Q$.  It follows that if the above matrix has orthonormal columns, then $Q^{*} P = 0$, and thus the frames $\{x_n\}$ and $\{y_n\}$ are orthogonal.

($\Rightarrow$) Conversely, suppose the Parseval frames are orthogonal.  Note that by Proposition 1, the left part $P$ of the above matrix has orthonormal columns; likewise the right part of the matrix $Q$ also has orthonormal columns.  By equation (\ref{E:ortho}), we must have that the columns of the left part of the matrix are orthogonal to the columns of the right part of the matrix.  Hence, the columns of the matrix form an orthonormal set.
\end{proof}

Note that if $\{x_n\}$ is orthogonal to $\{y_n\}$, then $\{y_n\}$ is orthogonal to $\{x_n\}$.

Let $\Bb{X} := \{x_n\}_{n=1}^{M} \subset H$; the analysis operator $\Theta_{\Bb{X}}$ of $\{x_n\}$ is given by:
$$ \Theta_{\Bb{X}}: H \to \Bb{F}^M : v \mapsto (\langle v, x_1 \rangle, \langle v, x_2 \rangle, \dots , \langle v, x_M \rangle ). $$
The matrix representation of $\Theta_{\Bb{X}}$ is given as the matrix $P$ in Proposition \ref{P:ortho}.  The proof of Proposition \ref{P:ortho} shows that two frames $\{x_n\}$ and $\{y_n\}$ are orthogonal if and only if their analysis operators $\Theta_{\Bb{X}}$ and $\Theta_{\Bb{Y}}$ have orthogonal ranges in $\Bb{F}^M$.

\subsection{Encryption Using Orthogonal Frames}

We present here an overview of our proposed private key encryption scheme using orthogonal frames.  For motivation, consider that the One-Time Pad is an unconditionally secure cipher, which is optimal of all unconditionally secure ciphers in terms of key length \cite{HAC}.  Our encryption scheme, which is similar to a subband coding scheme, is an effort to approximate the One-Time Pad.  The (private) key for this encryption scheme is two orthogonal Parseval frames $\{x_n\}_{n=1}^{M} \subset H$ and $\{y_n\}_{n=1}^{M} \subset K$.  Let $\Theta_{\Bb{X}}$ and $\Theta_{\Bb{Y}}$ respectively denote their analysis operators.  Suppose $m \in H$ is a message; let $g \in K$ be a non-zero vector chosen at random.  The ciphertext $c \in \Bb{F}^{M}$ is given as follows:
$$ c := \Theta_{\Bb{X}} m + \Theta_{\Bb{Y}} g.$$
To recover the message, we apply $\Theta_{\Bb{X}}^{*}$:
\begin{align*}
\Theta_{\Bb{X}}^{*} c &= \Theta_{\Bb{X}}^{*} \Theta_{\Bb{X}} m + \Theta_{\Bb{X}}^{*} \Theta_{\Bb{Y}} g \\
&= \sum_{n = 1}^{M} \langle m, x_n \rangle x_n + \sum_{n = 1}^{M} \langle m, y_n \rangle x_n \\
&= m + 0
= m.
\end{align*}

There are several things to note about our scheme:
\begin{enumerate}
\item The frame $\{x_n\}$ need not be Parseval, but Parseval frames are in general easier to work with.  Since the Parseval frames form only a small subset of all possible frames, using general frames would allow a much greater choice of specific encryption keys.
\item The frame $\{y_n\}$ need not be Parseval; it need not even be a frame, though again Parseval frames simplify matters.  If $\{y_n\}$ is not a frame, then $\Theta_{\Bb{Y}}$ has non-trivial kernel, and $\Theta_{\Bb{Y}}g$ could be 0 if g is chosen to be in the kernel.  (Below we will actually use scalar multiples of Parseval frames for both $\{x_n\}$ and $\{y_n\}$.)
\item Just as with the One-Time Pad, when done properly, encoding a message twice results in two different ciphertexts.
\item Unlike the One-Time Pad, in which a brute force attack results in all possible plaintexts, it appears unlikely that a brute force attack on our system would result in the same.  Our simulations indicate that an attack produces either a text which is very close to the original plaintext or is gibberish (see graphs below for more.)  However, at this time, we cannot prove why this is so.
\end{enumerate}

\begin{proposition}
If $\{x_n\}_{n=1}^{M} \subset H$ and $\{y_n\}_{n=1}^{M} \subset K$ are orthogonal frames, then $M \geq dim(H) + dim(K)$.
\end{proposition}
\begin{proof}
Let $\Theta_{\Bb{X}}$ and $\Theta_{\Bb{Y}}$ be the respective analysis operators.  Note that by the (lower) frame inequality in equation \ref{E:frame}, both $\Theta_{\Bb{X}}$ and $\Theta_{\Bb{Y}}$ are one-to-one.  Moreover, the orthogonality of the frames is equivalent to the orthogonality of the ranges of $\Theta_{\Bb{X}}$ and $\Theta_{\Bb{Y}}$.  Combining these two observations establishes the proposition.
\end{proof}

For convenience, we will assume that $M = dim(H) + dim(K)$.  The ciphertext is
$$ c = \Theta_{\Bb{X}} m + \Theta_{\Bb{Y}} g $$
where $\{x_n\}$ and $\{y_n\}$ are orthogonal Parseval frames.  Since they are orthogonal, we write
$$ c = \left( \Theta_{\Bb{X}} | \Theta_{\Bb{Y}} \right) m \oplus g $$
where the matrix $\left( \Theta_{\Bb{X}} | \Theta_{\Bb{Y}} \right)$ is an isometry.  Therefore, our encryption procedure involves generating a large orthogonal matrix.

The next section discusses several ways of constructing such matrices.  Since the encryption scheme is a private key system, we wish to have a relatively small key size; that is to say that the entire matrix is too much information to be used as the key.  We discuss below some of the strengths and weaknesses of the various construction techniques.



\section{Five Encryption Schemes} \label{S:schemes}

The cipher algorithm depends upon generating a pair of random orthogonal frames, each of which is the size of the message.  This is equivalent to producing a random orthogonal matrix of twice the size of the message.  We investigate here several methods for doing so.  The first method takes the view of producing orthogonal frames using Fourier frames.  The remaining methods take the view of producing orthogonal matrices.

Once the orthogonal frames, or orthogonal matrix, is determined, the encryption and decryption process is the same.  If the frames are given by $\Bb{X}$ and $\Bb{Y}$, then we write the matrix $(\Theta_{\Bb{X}} | \Theta_{\Bb{Y}} )$; if on the other hand the matrix is $A$, we think of $A = (\Theta_{\Bb{X}} | \Theta_{\Bb{Y}})$.  Given a message $m$, choose at random a vector $g$, called the ``garbage'' or ``noise'', and compute $(\Theta_{\Bb{X}} | \Theta_{\Bb{Y}} ) m \oplus g = c$ to yield the cipher text $c$.  The recipient computes 
$$ (\Theta_{\Bb{X}} | \Theta_{\Bb{Y}} )^{T} c = (\Theta_{\Bb{X}} | \Theta_{\Bb{Y}} )^{T} (\Theta_{\Bb{X}} | \Theta_{\Bb{Y}} ) m \oplus g = K m \oplus 0 = K m, $$
where $K$ is the square of the norm of any column of the matrix $\Theta_{\Bb{X}}$.  Dividing by $K$ then reproduces the message.

\subsection{Scheme \#1}

The first algorithm utilizes the Discrete Cosine Transform.  The original idea came from using the Discrete Fourier transform, which involves complex exponentials.  The Discrete Cosine Transform, in matrix form, is given by:
$$ C=[c_{kn}]=\left[ \lambda_k\sqrt{\frac{2}{M}} \cos \left\{ \frac{k\pi}{M}(n+1/2) \right\} \right], $$
where $n=1,\dots, M$, $\lambda_1=1/\sqrt{2}$ and $\lambda_k=1$ for all $k=2,\dots, N$.  Note that this is normalized to be a unitary matrix.  Assuming that $M=2N$, one can permute the columns of $C$ to yield $C'$, and divide the resulting matrix in half vertically:
$$ C' = (\Theta_{\Bb{X}} | \Theta_{\Bb{Y}} ). $$
The resulting divided matrix can then be viewed as the analysis operators for two orthogonal frames, each for $\Bb{R}^N$, consisting of cosine bases projected onto smaller subspaces, (Proposition \ref{P:ortho}, see also \cite{ALTW02}).  Moreover, the frame vectors can be weighted, which is accomplished by a diagonal, invertible matrix $D$.  Let $P$ denote a permutation matrix.

The (private) key for the cipher then consists of the matrix $D$ (or simply its diagonal entries), and the permutation corresponding to $P$.  The encryption algorithm of a message $m$ of length $N$ then consists of randomly generating a garbage vector $g \in \Bb{R}^N$ and computing the ciphertext $c$:
$$ c = C D P (m \oplus g). $$
To decrypt the message, we apply the matrix $Q P^{T}D^{-1}C^T$ to the ciphertext, where $Q$ is the projection of $\Bb{R}^M$ onto the first $N$ co-ordinates:
$$Q P^{T}D^{-1}C^T C D P(m \oplus g) = Q (m \oplus g) = m. $$

\begin{remark}
We note that the only knowledge unknown to an adversary is $D$ and $P$; the adversary will know $C$.  Hence, $C$ is irrelevant to the cipher algorithm.  Because of this, the algorithm reduces to rearrangement followed by weighting of the entries of the message and the garbage.  We conclude that our first algorithm is a poor one.
\end{remark}

\subsection{Scheme \#2}

The second scheme involves using Hadamard arrays to generate orthogonal matrices.  We first start with the definition of Hadamard arrays.  We remark here that this scheme is related to linear codes \cite{delsarte}.
\begin{definition} \cite{Wallis} A Hadamard array $H[h,k,\lambda ]$ based on the indeterminates $x_1,~x_2,\ldots,x_k$, with $k\leq h$, is an $h\times h$ matrix with entries chosen from $\{\pm x_1,~\pm x_2,\ldots,\pm x_k\}$ in such a way that:
\begin{enumerate}
\item In any row there are $\lambda$ entries $\pm x_1$, $\lambda$ entries $\pm x_2$, $\ldots,$ $\lambda$ entries $\pm x_k$, and similarly for the columns. 
\item The rows and columns are (formally) pairwise orthogonal, respectively.
\end{enumerate}
\end{definition}

The matrices we use for our encryption scheme are of $h=k$, $\lambda=1$.  The only possible Hadamard arrays of this type are for $h=1,2,4,8$ \cite{ag}.  For indeterminants $A$ through $H$, we have the Hadamard array
$$H[8,8,1]=\begin{bmatrix}A&B&C&D&|&E&F&G&H\\-B&A&D&-C&|&F&-E&-H&G\\-C&-D&A&B&|&G&H&-E&-F\\-D&C&-B&A&|&H&-G&F&-E\\
-E&-F&-G&-H&|&A&B&C&D\\-F&E&-H&G&|&-B&A&-D&C\\-G&H&E&-F&|&-C&D&A&-B\\-H&-G&F&E&|&-D&-C&B&A\\
\end{bmatrix}.$$

For $\Theta=H[8,8,1]$, $\Theta^T \Theta=K I_8$ where $K=A^2+B^2+\dots + H^2$.

The Hadamard arrays allow easy construction of matrices (and hence tight frames) needed in our encryption schemes.  For the encryption process, we now have only $\Theta$ to construct instead of computing the matrices $C$, $D$, and $P$.

The encryption process starts with a message $m$ of arbitrary length, and dividing $m$ into blocks $m_1,\dots , m_q$ of length 4 (padding the last block with $0$'s if necessary).  Then random vectors $g_1, \dots , g_q$ of length 4 are chosen, and the matrix $N$ is applied successively to $m_i \oplus g_i$.  The ciphertext is then 
$$ c = \Theta(m_1 \oplus g_1) \oplus \dots \oplus \Theta(m_q \oplus g_q) . $$

The message is then decrypted by dividing $c$ into blocks $c_1, \dots, c_q$ of size 8, computing $K \Theta^{T} c_i$ for $i=1, \dots ,q$, and reconstructing the message using the first four entries of these resulting blocks.

\begin{remark}
Because of the ease of construction of the Hadamard arrays, the system is quite easy to implement.  Unlike the first scheme, the key for the recipient has now been reduced to knowing the chosen entries for $\Theta$, hence in this case the key is the entries $A,B, \dots, H$ of the matrix $\Theta$.  Since Hadamard arrays are small, however, we wish to find an algorithm to generate larger orthogonal matrices.
\end{remark}


\subsection{Scheme \# 3}

Our next scheme is an attempt to produce larger orthogonal matrices.  Starting with Hadamard arrays $A$ and $M$ with $A^TA=kI_8$ and $M^TM=pI_8$ for constants $k$ and $p$, we construct a new $16 \times 16$ orthogonal matrix 
$$S = \begin{bmatrix}A&MA\\-M^TA&A\end{bmatrix}.$$
Repeat this procedure with Hadamard arrays $B$ and $N$ to get
$$T = \begin{bmatrix}B&NB\\-N^T B&B\end{bmatrix}.$$
The matrices $S$ and $T$ are then used to construct a $32 \times 32$ orthogonal matrix:
$$U = \begin{bmatrix}S&TS\\-T^T S&S\end{bmatrix}.$$
This ``blow up'' construction is iterated to get the appropriate size matrix for our plain text.

\begin{remark}
In this encryption scheme, the key is the entries of the matrices $A$, $B$, $M$, $N$, etc., and their positions in the construction.  This method, however is computationally inefficient.
\end{remark}

\subsection{Scheme \#4}

We first define the tensor product, $\otimes$, of two matrices, $A$ and $B$.  The sizes of the matrices is irrelevant.
\begin{definition}\cite{comb} Let $$A=\begin{bmatrix}a_{11}&a_{12}&\cdots&a_{1n}\\
            \vdots&~      &\ddots&~     \\
            a_{m1}&a_{m2}&\cdots&a_{mn}
    \end{bmatrix}.$$
Then $$A\otimes B:=\begin{bmatrix}a_{11}B&a_{12}B&\cdots&a_{1n}B\\
            \vdots&~      &\ddots&~     \\
            a_{m1}B&a_{m2}B&\cdots&a_{mn}B
    \end{bmatrix}.$$
\end{definition}

If $A$ is an $m\times n$ and $B$ is a $p\times q$, then $A\otimes B$ is an $mp\times nq$ matrix.  The tensor product will be the critical element of construction in this and the next scheme.  Note that if $A$ and $B$ are orthogonal matrices, then $A \otimes B$ is also an orthogonal matrix.

\begin{definition}
A Hadamard matrix is a square orthogonal matrix with entries consisting of $\pm 1$'s.
\end{definition}

We start with an Hadamard \textit{matrix} (not an array), $H$, of a chosen size $2^p$, and then two Hadamard \textit{arrays}, $A$ and $B$ of choice sizes 2,4, or 8.  We then construct the new matrix via the tensor products:
$$C=\begin{bmatrix}H\otimes A & (H\otimes B)(H\otimes A)\\ -(H\otimes B)^T(H\otimes A)& H\otimes A\end{bmatrix}.$$
$C$ is now an orthogonal matrix.  This matrix is size adaptive with respect to powers of 2 since each matrix is of some order of 2, and the size of $H$ can be chosen.

However, the Hadamard matrix property that $H^TH=I_n$ is actually a disadvantage.  Let
$$H=\begin{bmatrix}1&1&1&1\\1&-1&1&-1\\1&1&-1&-1\\1&-1&-1&1\end{bmatrix}.$$  
Then our matrix is
$$C=\begin{bmatrix}H\otimes A & (H\otimes B)(H\otimes A)\\ -(H\otimes B)^T(H\otimes A)& H\otimes A\end{bmatrix}=$$
$$\begin{bmatrix}   A&A&A&A &|& 4BA&0&0&0 \\
            A&-A&A&-A &|& 0&4BA&0&0 \\
            A&A&-A&-A &|& 0&0&4BA&0 \\
            A&-A&-A&A &|& 0&0&0&4BA \\
            -&-&-&-   &|& -&-&-&-   \\
            -4B^{T}A&0&0&0 &|& A&A&A&A   \\
            0&-4B^{T}A&0&0 &|& A&-A&A&-A \\
            0&0&-4B^{T}A&0 &|& A&A&-A&-A \\
            0&0&0&-4B^{T}A &|& A&-A&-A&A \end{bmatrix}.$$

The resulting matrix is relatively sparse, which is undesirable for maintaining secrecy.

\subsection{Scheme \#5}

We choose $p$ Hadamard arrays $H_1,H_2,\ldots,H_p$.  Each array can have its own size, say $e_i\times e_i$ for $1\leq i\leq p$, where each $e_i$ is either 2,4, or 8.  We then construct our $e_1e_2\cdots e_p$-sized matrix $M$ by the tensor product of these $p$ matrices:
$$M = \bigotimes^{p}_{i=1}H_i:=H_1\otimes H_2\otimes\cdots\otimes H_p.$$
The ciphertext then is $c = M(m\oplus g)$.  With this construction, we eliminate the sparsity that was shown in scheme \#4.  Note that the key in this case is the entries of the first rows of $H_1$ to $H_p$, hence is an array of numbers of size $e_1 + e_2 + \cdots + e_p$, and hence is relatively small.


We ran some numerical experiments, using scheme \#5 to obtain information regarding several things:
\begin{enumerate}
\item  We wanted to see if a brute force attack would be a feasible way of defeating the cipher.  The results of the experiments and also the computations below suggest that the answer is no.
\item  One advantage of the One Time Pad is that a brute force attack results in all possible plaintext messages, forcing an adversary to choose which was the original message.  We wanted to determine if this was also true of our proposed cipher.  The results of our experiments indicate that the answer to this is also no.
\item  Finally, we wanted to determine if the size of the entries of the garbage vector $g$ mattered.  The experiments and the computations below suggest that the answer is yes.
\end{enumerate}
The results of our experiments, in the form of graphs, are given below.

\section{Experimental Results and Conclusions} \label{S:results}

We want to know how accurate a guess has to be in order to break the cipher.  We suppose that an adversary knows that we are using scheme \#5, that is the adversary knows the structure of the matrix $M$, but not the entries.  We let $M$ be the original matrix of size $n$, $\tilde{M}$ be the adversary's guess, and $w$ be the original plaintext $m$ concatenated with the garbage $g$ (i.e. $w=m \oplus g$). Then we consider $\tilde{w} := (1/\tilde{k})\tilde{M}^TMw$ where $\tilde{k} = \|\tilde{M}\|^2$.  Since we assume that the structure of $M$ is known by the adversary, we consider $\tilde{M}=M+P$, where $P$ is a matrix with the same structure as $M$.  For simplicity, we let $M_i$ denote the $i$th row of the matrix $M$ and likewise for $P$. Note that $k=\langle M_i,M_i\rangle$ since $(1/k)M^TM=I_n$, and $\tilde{k}=\langle\tilde{M}_i,\tilde{M}_i\rangle=\langle M_i+P_i,M_i+P_i\rangle=||M_i||^2+2\langle M_i,P_i\rangle+||P_i||^2$.

We rewrite to get the following:
\begin{multline*}
(1/\tilde{k})\tilde{M}^TM=(k/\tilde{k})I+(1/\tilde{k})P^TM= \\
\bfr{\langle M_i,M_i\rangle}{\langle M_i+P_i,M_i+P_i\rangle}I+\bfr{1}{\langle M_i+P_i,M_i+P_i\rangle}\begin{bmatrix}\langle P_1^T,M^T_1\rangle & \cdots & \langle P_1^T,M^T_n\rangle \\ \vdots & \ddots & ~ \\ \langle P_n^T,M^T_1\rangle & \cdots & \langle P_n^T,M^T_n\rangle \end{bmatrix}.
\end{multline*}

Let $\tilde{w}=\left(\tilde{w}_1,\tilde{w}_2,\ldots,\tilde{w}_n\right).$  Then we have that for $1 \leq j \leq n$:
$$\tilde{w}_j = \left(\bfr{\langle P_j^T,M_j^T\rangle+\langle M_j,M_j\rangle}{\langle M_j+P_j,M_j+P_j\rangle}w_j+\sum^n_{\begin{array}{c}i=1 \\ i\not=j\end{array}}\bfr{\langle P_j^T,M_i^T\rangle}{\langle M_i+P_i,M_i+P_i\rangle}w_i\right)$$

For an adversary's guess to be close, $$\bfr{\langle P_j^T,M_j^T\rangle+\langle M_j,M_j\rangle}{\langle M_j+P_j,M_j+P_j\rangle} \approx 1$$ and $$\sum^n_{\begin{array}{c}i=1 \\ i\not=j\end{array}}\bfr{\langle P_j^T,M_i^T\rangle}{\langle M_i+P_i,M_i+P_i\rangle} \approx 0.$$
We break this up into cases.
\begin{itemize}
\item[Case 1:] Assume $||P||$ is relatively large compared to $\|M\|$; that is, the guess is far from the actual matrix.  We have
\begin{multline*}
\left|\bfr{\langle P_i^T,M_i^T\rangle}{\langle M_i+P_i,M_i+P_i\rangle}\right|=\left|\bfr{\langle P_i^T,M_i^T\rangle}{||M_i||^2+2\langle P_i,M_i\rangle+||P_i||^2}\right|= \\
\left|\bfr{\langle P_i^T,M_i^T\rangle/||P_i||^2}{(||M_i||^2/||P_i||^2)+(2\langle P_i,M_i\rangle/||P_i||^2)+1}\right|\to 0 \mbox{ as } ||P||\to\infty.
\end{multline*}
However, when we look at the $\tilde{w}_i$ coefficients, we see the following:
$$\left|\bfr{\langle M_j,M_j\rangle}{\langle M_j+P_j,M_j+P_j\rangle}\right|= \left|\bfr{(\langle M_j,M_j\rangle/||P_j||^2)}{(||M_i||^2/||P_i||^2)+(2\langle P_i,M_i\rangle/
||P_i||^2)+1}\right|\to 0 \mbox{ as } ||P_j||\to\infty.$$

\item[Case 2:] We assume $||P||$ is small relative to $||M||$; that is, the guess is close. Then we have using the same arguments:
\begin{multline*}
\left|\bfr{\langle P_i^T,M_i\rangle}{\langle
M_i+P_i,M_i+P_i\rangle}\right|=\left|\bfr{\langle
P_i^T,M_i\rangle}{||M_i||^2+2\langle M_i,P_i\rangle
+||P_i||^2}\right|= \\
\left|\bfr{\langle P_i^T,M_i\rangle/||M_i||^2}{1+(2\langle
M_i,P_i\rangle/||M_i||^2)+(||P_i||^2/||M_i||^2)}\right|\to 0 \mbox{
as }||P_i||\to 0.
\end{multline*}
So, the better the guess, the smaller the `extra' coefficients will be. Likewise, for the $\tilde{w}_j$ coefficients,
\begin{multline*}
\left|\bfr{\langle M_j,M_j\rangle}{\langle
M_j,P_j\rangle}\right|=\left|\bfr{||M_j||^2}{||M_j||^2+2\langle
M_j,P_j\rangle +||P_j||^2}\right|= \\
\left|\bfr{1}{1+(2\langle
M_j,P_j\rangle/||M_j||^2)+(||P_j||^2/||M_j||^2)}\right|\to 1
\mbox{ as }||P_j||\to 0.
\end{multline*}

\end{itemize}

Our first question is whether an adversary can figure out how small the perturbation $P$ must be in order to get a "good guess".  The adversary knows the size of $M$ and $||Mw||$; we assume additionally that the adversary knows the structure of $M$.  For convenience, assume that the encryption matrix $M=A\otimes B\otimes C$ for 3 Hadamard arrays, $A,B$, and $C$.  We then let $\tilde{M}=(A+a)\otimes(B+b)\otimes(C+c)$ for (small norm) perturbation matrices $a,b$ and $c$.  We reformulate our question: How big can $||a||,||b||$, and $||c||$ be such that $||M^TMw-\tilde{M}^TMw||<\ep$, where $\ep$ is some acceptable tolerance for error?  (Here, for a matrix $A$, $\| A \|$ denotes the operator norm of $A$.  Below, $\| \cdot \|$ shall denote both Hilbert space norm for vectors and operator norm for matrices.)

We let $||a||\approx||b||\approx||c||\approx\beta$ and $||A||\approx||B||\approx||C||\approx\gamma$.  We may assume that $\gamma\gg\beta$.  If we write out $\tilde{M}$ in terms of the tensor products, we get
$$\tilde{M}=A\otimes B\otimes C+A\otimes B\otimes c +\cdots+
a\otimes b\otimes c \mbox{ and } ||\tilde{M}||\leq\gamma^3+3\gamma^2\beta+3\gamma\beta^2+\beta^3.$$
Given any $\ep>0$, we choose $\delta=\ep/||Mw||$.  If $|3\gamma^2\beta|<\delta$, then
\begin{align*}
||M^TMw-\tilde{M}^TMw|| &\leq||M^T-\tilde{M}^T|| ||Mw|| \leq (3\gamma^2\beta+3\gamma\beta^2+\beta^3) ||Mw|| \\
& \approx 3(\gamma^2\beta) ||Mw|| <\delta ||Mw||=\ep.
\end{align*}

These computations suggest that the larger the entries of the garbage vector $g$ are, the closer a guess must be in order to reasonably recover the message.  This is corroborated by the experiments we ran (see the graphs below).  Thus, we can control the accuracy an adversary would need in order to break the cipher.

\subsection{Chosen-Plaintext Attack}

We will demonstrate here a chosen-plaintext attack on the cipher which will break the system.  A chosen-plaintext attack is an attack mounted by an adversary which chooses a plaintext and is then given the corresponding ciphertext.
\begin{theorem}
The encryption algorithm proposed above is vulnerable to a chosen-plaintext attack.
\end{theorem}
\begin{proof}
We assume the adversary knows the length of the message band and subsequently the length of the noise band.  Let the length of the message band be $N_m$ and the length of the noise band be $N_n$.  The attack is as follows:
\begin{enumerate}
\item[Step 1.] Determine the range of the noise band $K$ of $\Theta$.  That is, determine $(\Theta_{\Bb{X}} | \Theta_{\Bb{Y}}) (0 \oplus \Bb{R}^{N_n})$.  Choose any plaintext $m$ of size $N_m$.  Encode the plaintext twice, with output, say, $e_0$ and $e_1$.  Compute $e_1 - e_0 = \Theta (m \oplus g_1) - \Theta (m \oplus g_0) = \Theta (0 \oplus g_1 - 0 \oplus g_0)$.  Notice that this yields a vector $f_1 = \Theta (0 \oplus g_1 - 0 \oplus g_0)$ in the range of the noise band of $\Theta$.  Encode the plaintext a third time, with output $e_2$, and compute $f_2 = e_2 - e_0$.  Compute $f_3, \dots , f_m$ until the collection $\{f_1, \dots , f_m \}$ contain a linearly independent subset of size $N_n$.  This determines the range of the noise band $K$ of $\Theta$.
\item[Step 2.] Determine the range of the message band $T$ of $\Theta$.  That is, determine what is $(\Theta_{\Bb{X}} | \Theta_{\Bb{Y}}) (\Bb{R}^{N_m} \oplus 0)$.  Choose any plaintext $m_1$ of size $N_m$; encode the plaintext, with output $e_1$; then project $e_1$ onto the orthogonal complement of $K$.  This yields a vector $x_1$ in $T$.  Choose another plaintext $m_2$ and repeat, yielding vector $x_2 \in T$.  Repeat until the collection $\{x_1, \dots , x_q\}$ contains a linearly independent subset of size $N_m$.  This set determines $T$.
\item[Step 3.] Determine the message part of $\Theta$.  That is, determine $\Theta_{\Bb{X}}$.  Suppose in Step 2, $\{m_1, \dots , m_{N_m} \}$ is such that $\{x_1, \dots , x_{N_m} \}$ is linearly independent.  If we write $\Theta = ( \Theta_{\Bb{X}} | \Theta_{\Bb{Y}} )$, then we now have the following system of equations:
$$ ( \Theta_{\Bb{X}} | \Theta_{\Bb{Y}} ) m_k \oplus 0 = x_k \text{ for $k = 1, \dots N_m$.} $$
Given this system of equations, now solve for $\Theta_{\Bb{X}}$.
\item[Step 4.] Unencode ciphertexts.
Given any ciphertext $e$, the adversary computes the following:
\begin{align*}
K^{-1} (\Theta_{\Bb{X}} | 0 )^T e &= K^{-1} (\Theta_{\Bb{X}} | 0)^T (\Theta_{\Bb{X}} | \Theta_{\Bb{Y}}) m \oplus g \\
&= K^{-1} \Theta_{\Bb{X}}^T \Theta_{\Bb{X}} m \\
&= m
\end{align*}
where $K$ is the square of the norm of any column of $\Theta_{\Bb{X}}$.
\end{enumerate}

\end{proof}

\subsection{Concluding Remarks}
The proposed cipher appears to be robust to brute force attacks, but is not robust against a chosen-plaintext attack.  We mention, however, that we do not know if the scheme is robust to a known-plaintext attack.  Moreover, this is a private symmetric key cipher; it would be desirable if this method could be altered to be used as a public key cipher.  We reiterate that the McEliece cipher is a public key system and is similar in flavor to the cipher presented here.

The ultimate downfall of the cipher is the linearity.  We suggest that perhaps there is possibly a way of introducing non-linearity into the algorithm to defeat a chosen-plaintext attack.  However, at this point, we know of no methods to accomplish this.

\section{Pseudo-Code}

\subsection{Encoder.cpp}
\begin{enumerate}
\item Calculate Matrix
    \begin{enumerate}
    \item Input the possible range of entries for A, B, C
    \item Make A, B, C either 4x4 or 8x8 Hadamard arrays with entries chosen randomly from the range (for simplicity, we are using the 4x4 Hadamard array)
    \item Compute tensor product $A \oplus B \oplus C$
    \end{enumerate}

\item Encode Message
    \begin{enumerate}
    \item Compute $m \oplus g$ by converting the message to ASCII and filling $g$ with random numbers
    \item Compute $(A \otimes B \otimes C) (m \oplus g)$
    \end{enumerate}

\end{enumerate}

\subsection{Hacker.cpp}
Hacker.cpp--this code attempts a brute force method on a cypher
text.

\begin{enumerate}
    \item Input min, max, range of key guesses
    \item Input ciphertext
    \item For all possible values of the twelve variables in use
        \begin{enumerate}
        \item Fill the matrices with the possible values
        \item Tensor matrices together
        \item Calculate possible text messages
        \item Output text to file for later examination
        \end{enumerate}
\end{enumerate}

\subsection{Analyzer.cpp}
This code takes the output of Hacker.cpp and calculates the
frequency of occurrence of every ASCII symbol.
\begin{enumerate}
    \item For each line of text, count number of appearances of each ASCII value
    \item Output information to text file
\end{enumerate}


\section*{Acknowledgements}
This work was done while all three authors were at the University of Wyoming, at which time the first author was an undergraduate student, and the third author was a graduate student.  The first and second authors were supported by NSF grant DMS-0308634.  The third author was supported by a Basic Research Grant from the University of Wyoming.

We thank Bryan Shader, Eric Moorhouse, and Cliff Bergman for helpful discussions.

\section{Graphs}

How to read the following graphs.  We carried out the following computations to simulate a brute force attack on the cipher:
\begin{enumerate}
\item for a sample plaintext, encode the plaintext using scheme \#5 making the following choices: approximate entry size for the matrices and approximate size for the garbage entries;
\item decode the ciphertext using every combination of key entry and key entry $\pm 1$;
\item converted the decoded ciphertext in the previous step to ASCII values;
\item counted the appearance of each value in the resulting combinations.
\end{enumerate}
The graphs represent the number of appearances within all possible key guesses from step 2 above.  The plaintext is given in the title of the graph; the ASCII values are the $x$-axis of the graph, and the approximate key sizes and garbage sizes are given in the graph captions.  

Note that in figures 3 and 7, the key size and garbage size are the same.  The graphs show that most of the characters that appear in the simulated brute force attack are those that are in the original message.

\twocolumn

\begin{figure}[ht]
  \scalebox{.4}{\includegraphics{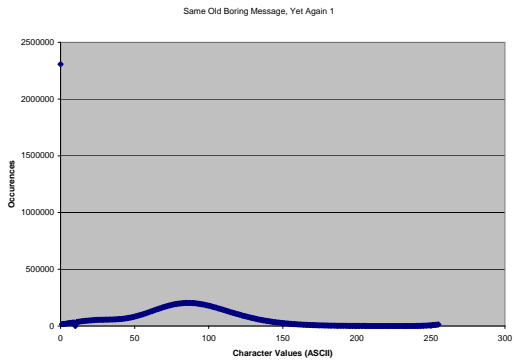}}
  \caption{Key: 5-7; Garbage: 128}
\end{figure}


\begin{figure}[ht]
  \scalebox{.4}{\includegraphics{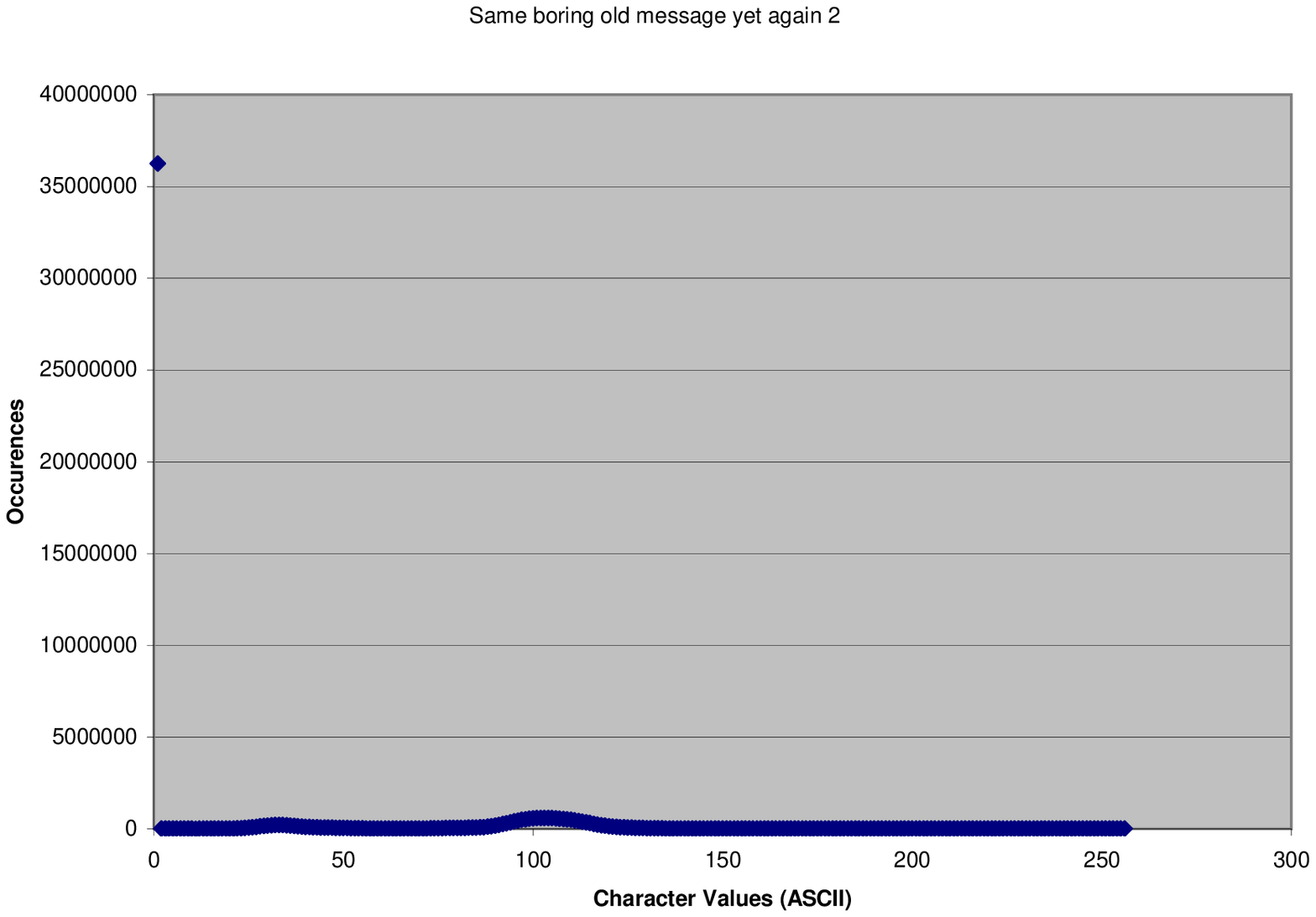}}
  \caption{Key: 25-27; Garbage: 128}
\end{figure}


\begin{figure}[ht]
  \scalebox{.4}{\includegraphics{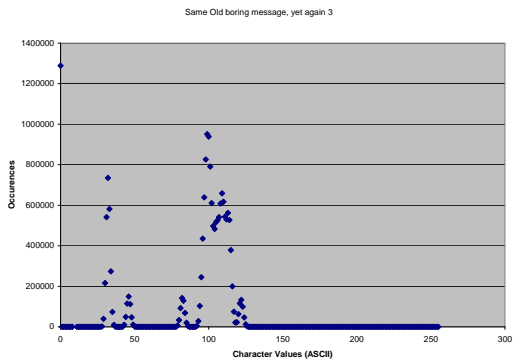}}
  \caption{Key: 100-102; Garbage: 128}
\end{figure}

\begin{figure}[ht]
  \scalebox{.4}{\includegraphics{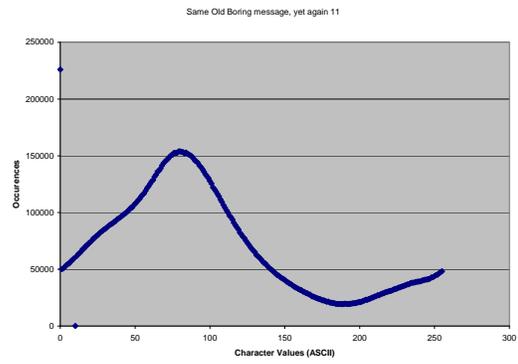}}
  \caption{Key: 5-7; Garbage: 1000}
\end{figure}

\begin{figure}[ht]
  \scalebox{.4}{\includegraphics{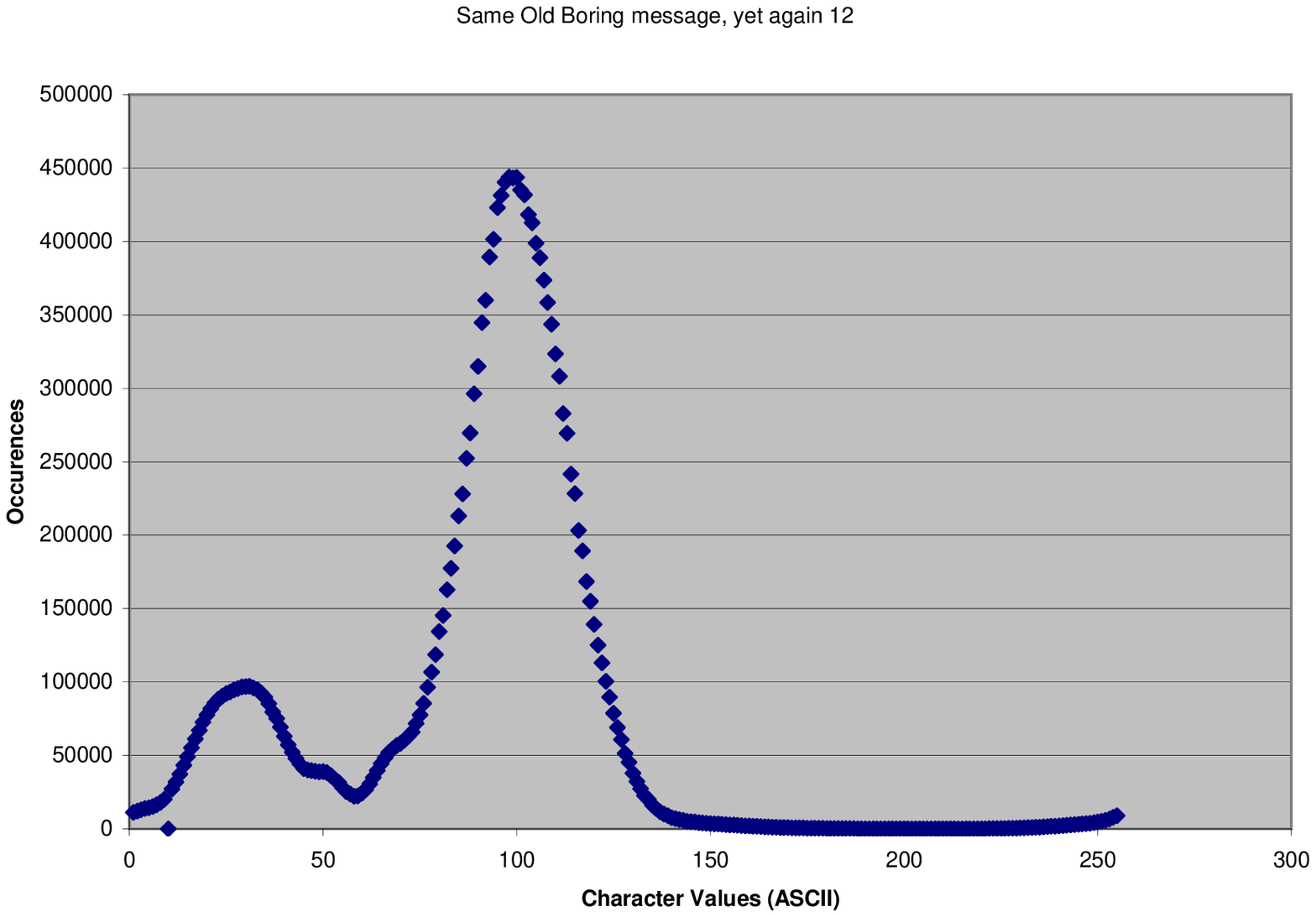}}
  \caption{Key: 25-27; Garbage: 1,000}
\end{figure}

\begin{figure}[ht]
  \scalebox{.4}{\includegraphics{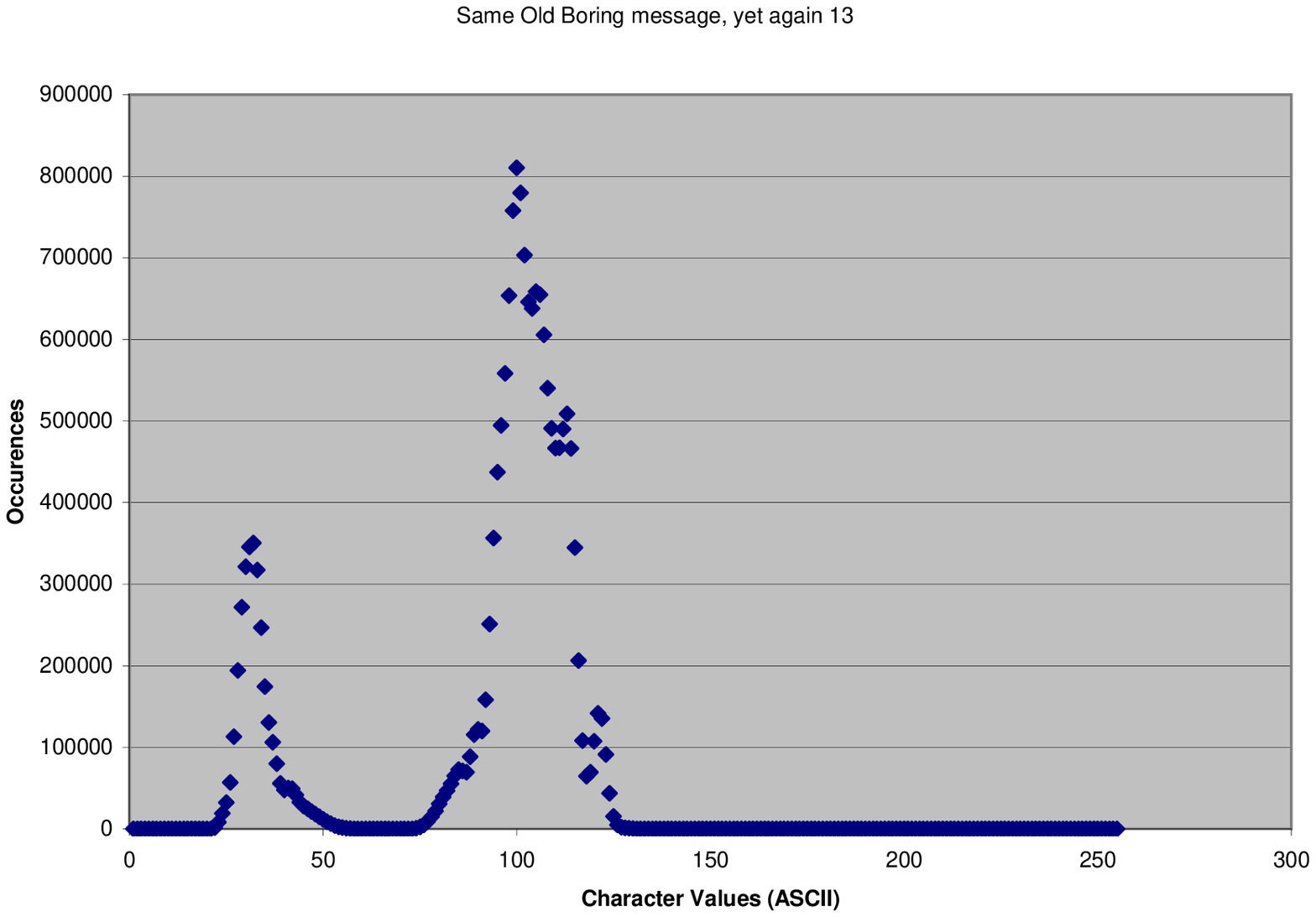}}
  \caption{Key: 100-102; Garbage: 1,000}
\end{figure}

\newpage

\begin{figure}[ht]
  \scalebox{.4}{\includegraphics{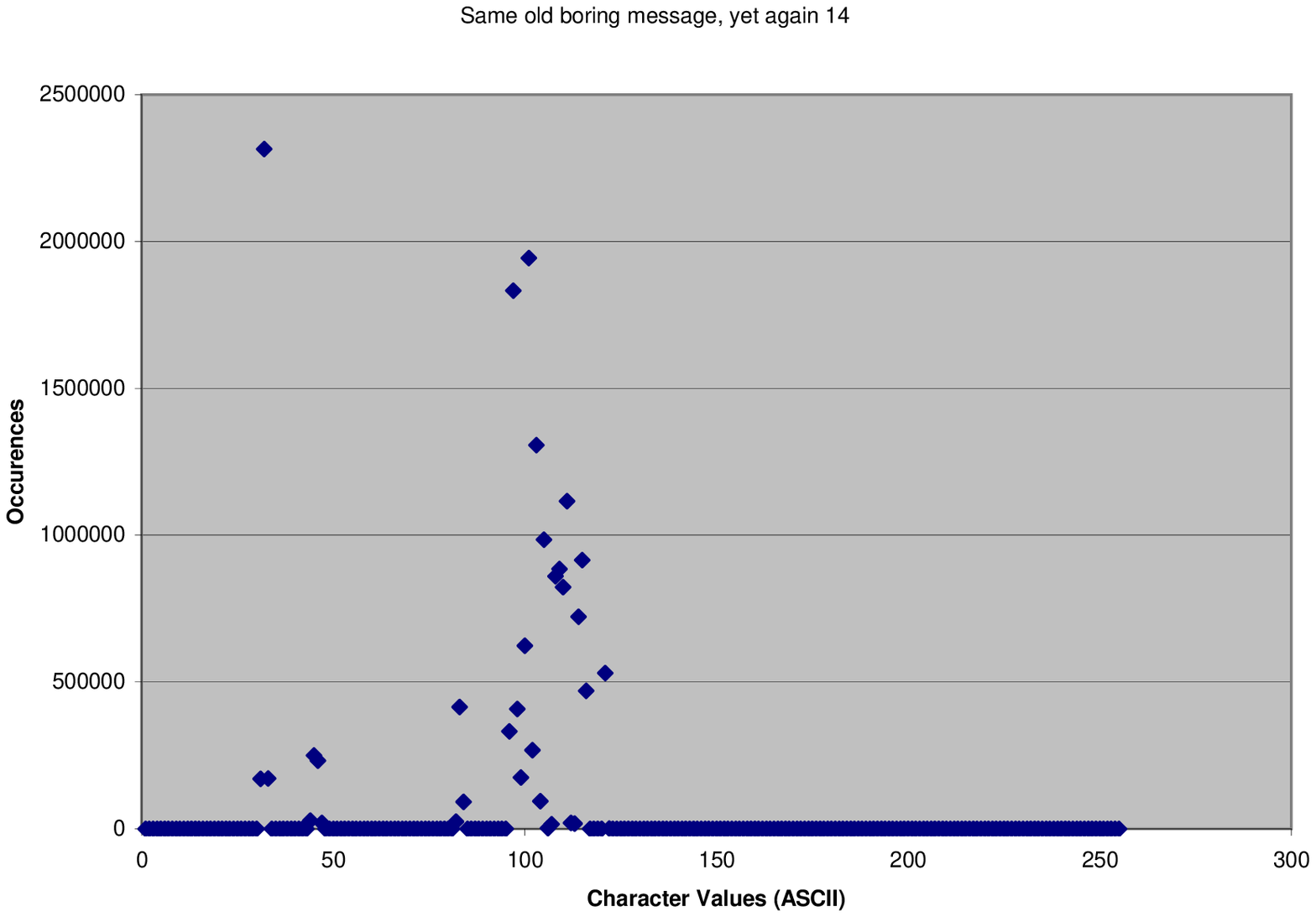}}
  \caption{Key: 1,000-1,002; Garbage: 1,000}
\end{figure}

\begin{figure}[ht]
  \scalebox{.4}{\includegraphics{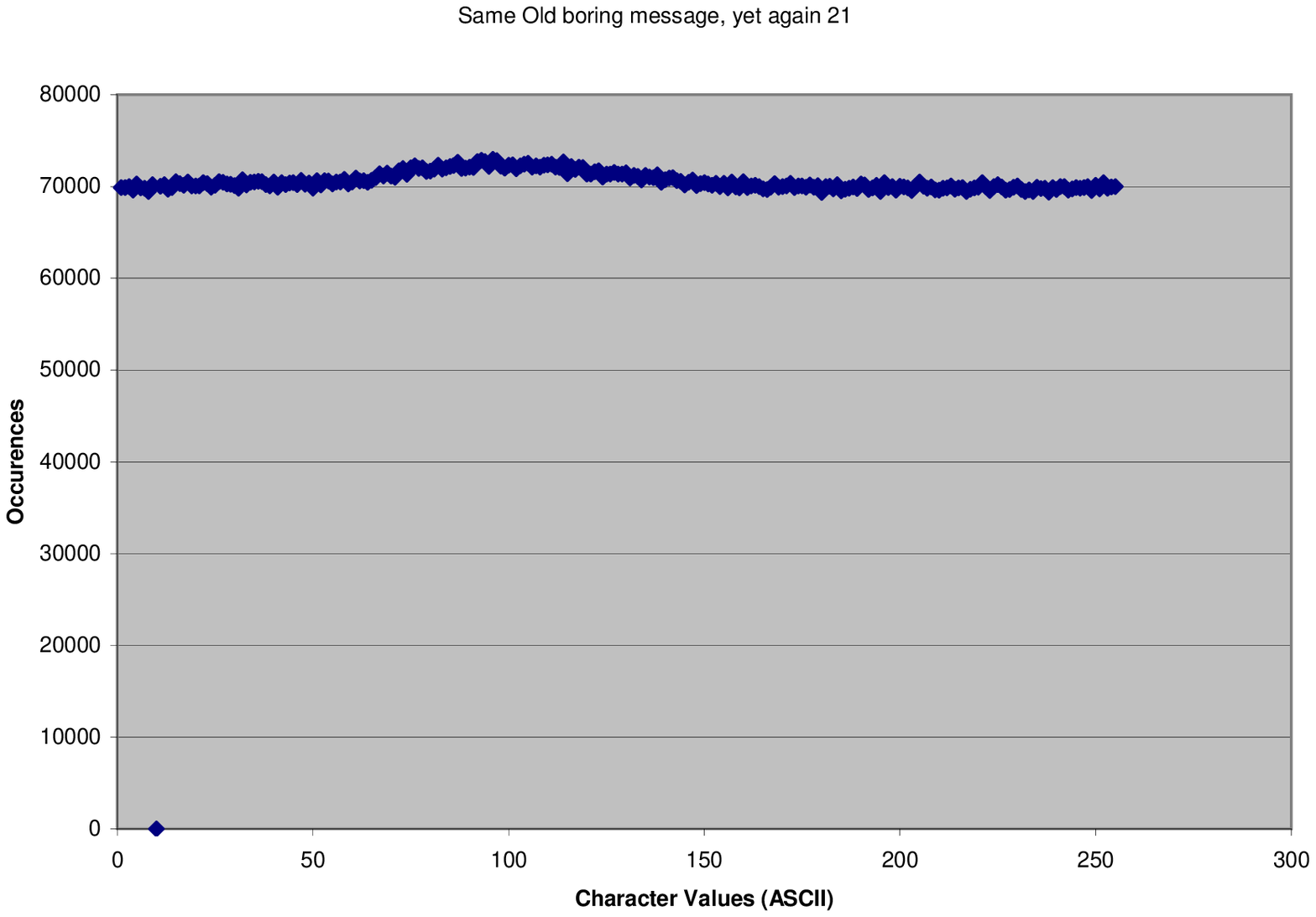}}
  \caption{Key: 5-7; Garbage: 100,000}
\end{figure}

\begin{figure}[ht]
  \scalebox{.4}{\includegraphics{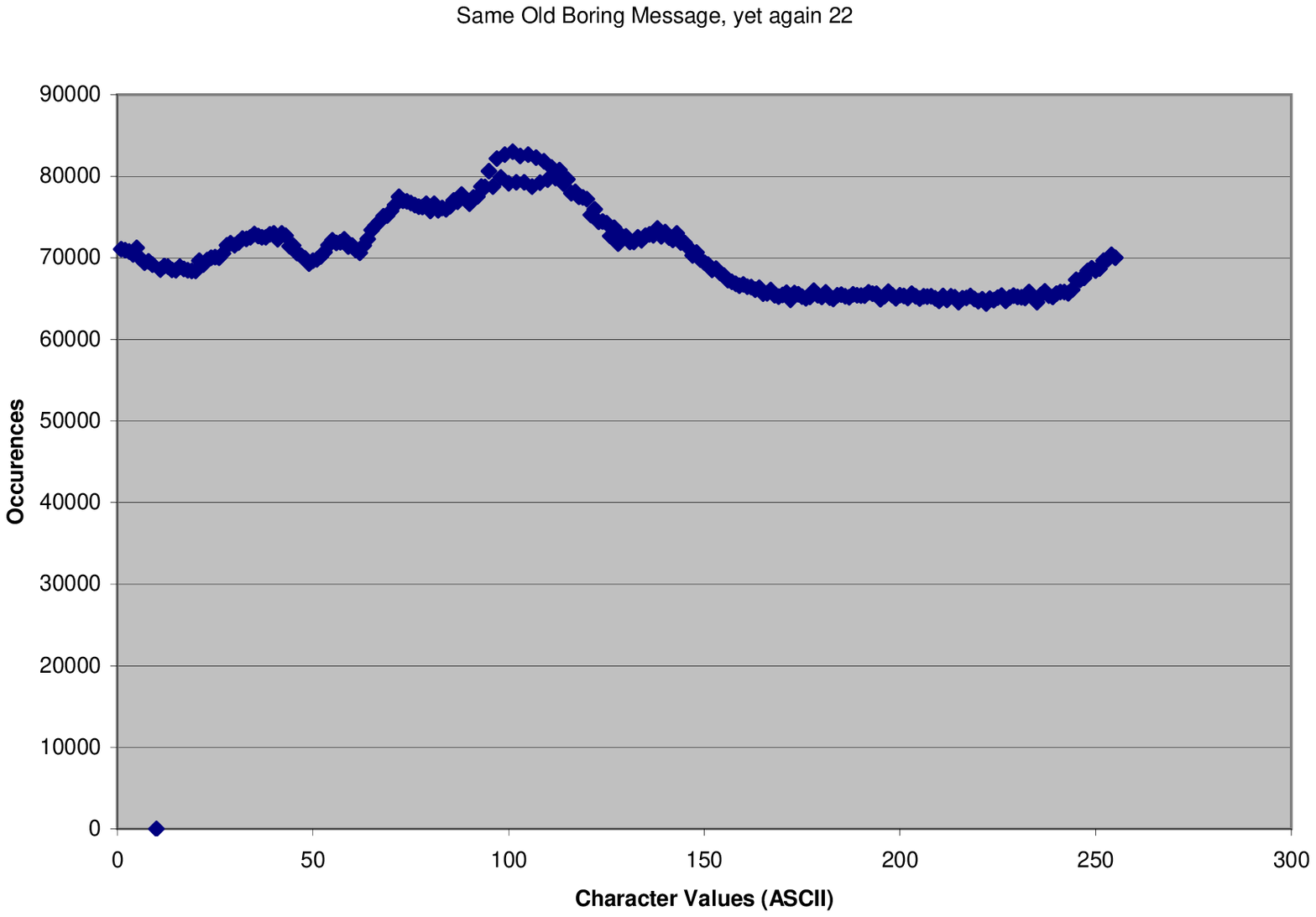}}
  \caption{Key: 25-27; Garbage: 100,000}
\end{figure}

\begin{figure}[ht]
  \scalebox{.4}{\includegraphics{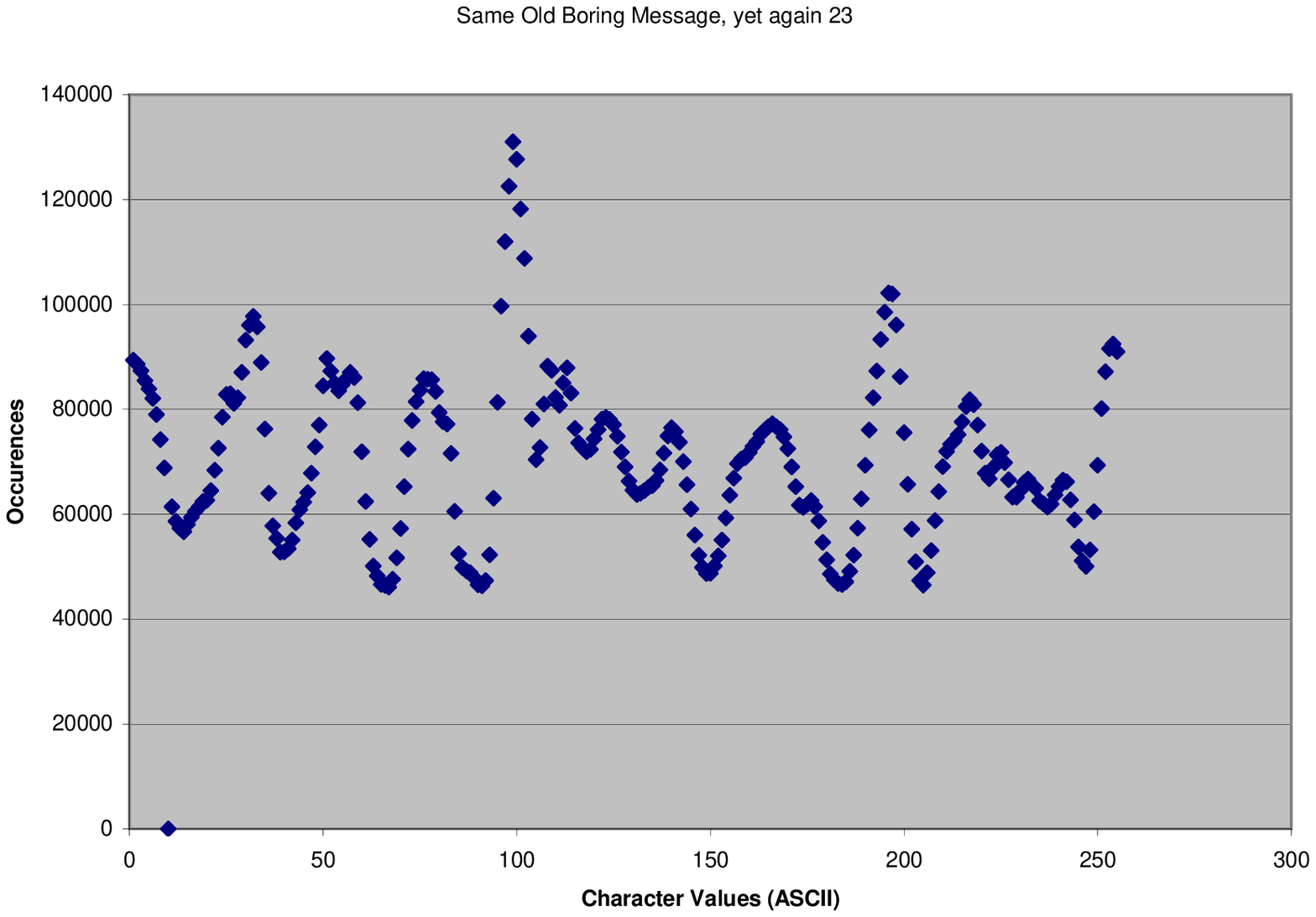}}
  \caption{Key: 100-102; Garbage: 100,000}
\end{figure}

\begin{figure}[ht]
  \scalebox{.4}{\includegraphics{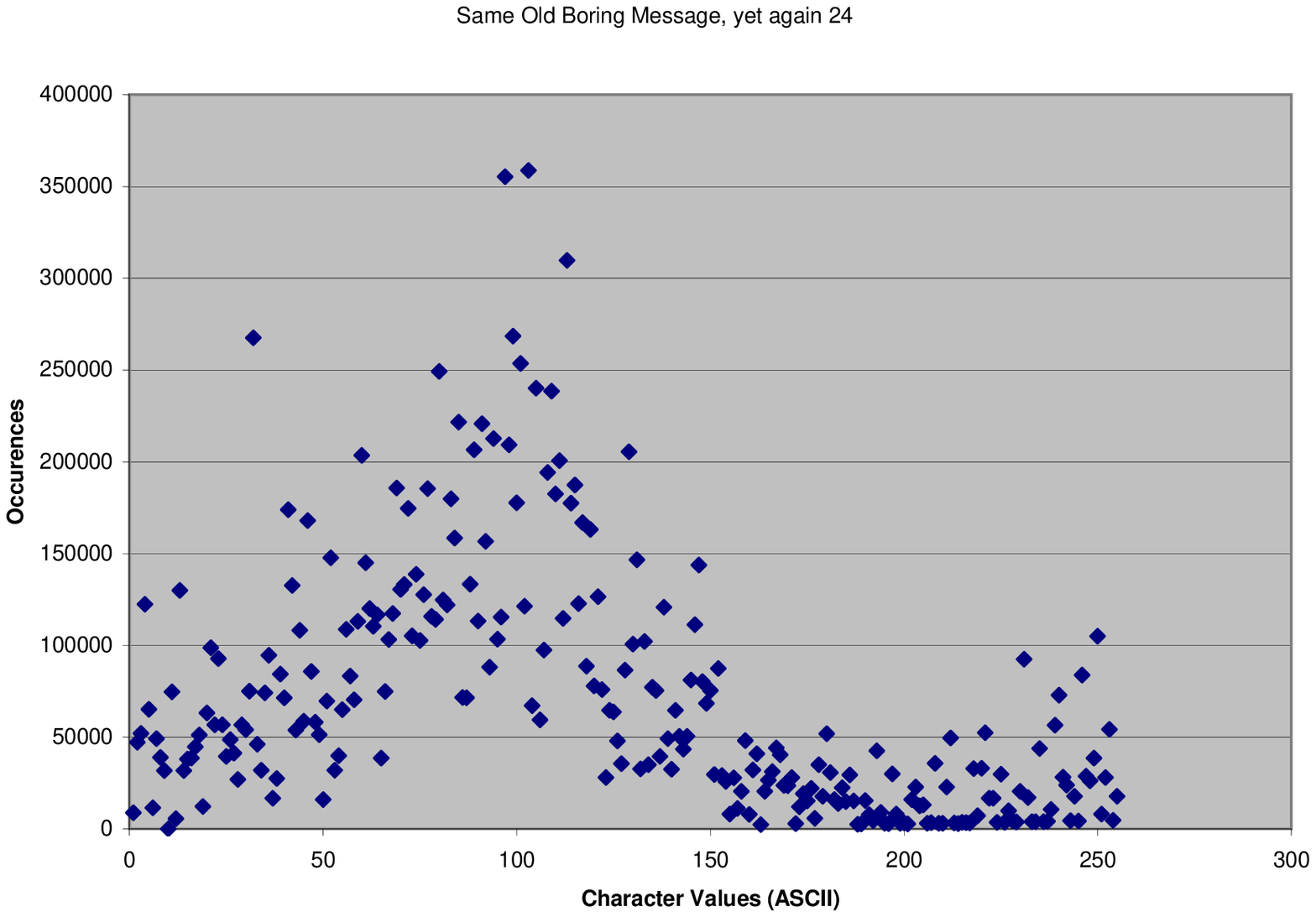}}
  \caption{Key: 1,000-1,002; Garbage: 100,000}
\end{figure}

\onecolumn





\end{document}